\documentclass[twocolumn]{aastex631}

\newcommand{\percentile}[1]{\ensuremath{P_{{#1}\%}}}
\newcommand{\kms}{\,km\,s$^{-1}$}
\newcommand{\masyr}{\,mas\,yr$^{-1}$}
\newcommand{\msun}{\,M$_\odot$}
\newcommand{\hst}{{\em HST}}
\newcommand{\gaia}{{\em Gaia}}
\newcommand{\swift}{{\em Swift}}
\newcommand{\magf}{\ensuremath{m_\textrm{\tiny F140W}}}
\newcommand{\sgr}{SGR\,1935+2154}
\newcommand{\snr}{SNR\,G57.2+08}
\newcommand{\flt}{\texttt{\_flt.fits}}
\newcommand{\pmra}{\ensuremath{\mu_{\alpha\star}}}
\newcommand{\pmdec}{\ensuremath{\mu_{\delta}}}
\newcommand{\pmrag}{\ensuremath{\mu_{\alpha\star, G}}}
\newcommand{\pmdecg}{\ensuremath{\mu_{\delta, G}}}
\newcommand{\pmtot}{\ensuremath{\mu}}
\newcommand{\vtot}{\ensuremath{\nu_T}}
\newcommand{\epochone}{2015.2}
\newcommand{\epochtwo}{2015.6}
\newcommand{\epochthree}{2016.4}
\newcommand{\epochfour}{2021.4}

\defcitealias{bedin2018}{BF18}

\submitjournal{ApJ}

\shorttitle{Long-term \hst\ monitoring of \sgr}
\shortauthors{Lyman et al.}
\graphicspath{{./}{figures/}}

\begin{document}

\title{The Fast Radio Burst-emitting magnetar \sgr{} -- proper motion and variability from long-term {\em Hubble Space Telescope} monitoring}

\author[0000-0002-3464-0642]{J. D. Lyman}
\affiliation{Department of Physics, University of Warwick, Coventry, CV4 7AL, UK}

\author[0000-0001-7821-9369]{A. J. Levan}
\affiliation{Department of Astrophysics/IMAPP, Radboud University Nijmegen, P.O. Box 9010, 6500 GL Nijmegen, The Netherlands}
\affiliation{Department of Physics, University of Warwick, Coventry, CV4 7AL, UK}

\author[0000-0002-9133-7957]{K. Wiersema}
\affiliation{Physics Department, Lancaster University, Lancaster, LA1 4YB, UK}
\affiliation{Department of Physics, University of Warwick, Coventry, CV4 7AL, UK}
\affiliation{School of Physics and Astronomy, University of Leicester, University Road, Leicester, LE1 7RH, UK}

\author{C. Kouveliotou}
\affiliation{Department of Physics, The George Washington University, Corcoran Hall, 725 21st St NW, Washington, DC 20052, USA}
\affiliation{GWU/Astronomy, Physics and Statistics Institute of Sciences (APSIS)}

\author[0000-0001-9842-6808]{A. A. Chrimes}
\affiliation{Department of Astrophysics/IMAPP, Radboud University Nijmegen, P.O. Box 9010, 6500 GL Nijmegen, The Netherlands}

\author{A. S. Fruchter}
\affiliation{Space Telescope Science Institute, 3700 San Martin Drive, Baltimore, MD 21218, USA}



\begin{abstract}

We present deep {\em Hubble Space Telescope} near-infrared (NIR) observations of the magnetar \sgr\ from June 2021, approximately 6 years after the first \hst\ observations, a year after the discovery of fast radio burst like emission from the source, and in a period of exceptional high frequency activity. Although not directly taken during a bursting period the counterpart is a factor of $\sim 1.5$ to $2.5$ brighter than seen at previous epochs with F140W(AB) = $24.65\pm0.02$\,mag. We do not detect significant variations of the NIR counterpart within the course of any one orbit (i.e. on minutes--hour timescales), and contemporaneous X-ray observations show \sgr\ to be at the quiescent level. With a time baseline of 6 years from the first identification of the counterpart we place stringent limits on the proper motion of the source, with a measured proper motion of $\mu = 3.1\pm1.5$\masyr. The direction of proper motion indicates an origin of \sgr\ very close to the geometric centre of \snr, further strengthening their association. At an adopted distance of $6.6\pm0.7$\,kpc, the corresponding tangential space velocity is $\vtot=97\pm48$\kms\ (corrected for differential Galactic rotation and peculiar Solar motion), although its formal statistical determination may be compromised owing to few epochs of observation. The current velocity estimate places it at the low end of the kick distribution for pulsars, and makes it among the lowest known magnetar kicks. When collating the few-magnetar kick constraints available, we find full consistency between the magnetar kick distribution and the much larger pulsar kick sample.

\end{abstract}



\section{Introduction} 
\label{sec:intro}

Magnetars are a diverse set of neutron stars with magnetic fields in excess of $B\sim10^{14}$\,G \citep[e.g.,][]{duncan1992, kouveliotou1998, kaspi2017}. In the Milky Way they manifest predominantly as soft-gamma repeaters (SGRs) and anomalous X-ray pulsars (AXPs), although some rotational powered pulsars (RRPs) have also exhibited magnetar-like behaviour \citep{gavriil2008}. The Galactic population currently numbers $>30$ objects \citep{olausenkaspi2014}, the majority of which have spin periods of 1 to 10 seconds, and characteristic spin-down ages of 100 to 10,000 years. 

Magnetars provide an ideal test-bed for several areas of extreme astrophysics since they probe matter under the effect of extreme density, magnetic field strength, and gravity.  In recent years, magnetars have acquired particular interest for their role as central engines 
in transient extragalactic events, including some of the most energetic events we know of. The creation and subsequent spin-down of new magnetars 
has been invoked to explain super luminous supernovae \citep[e.g.][]{kasen2010, woosley2010, inserra2013, dessart2019}, and both long- and short-duration Gamma-Ray Bursts (GRBs) \citep[e.g.][]{usov1992, zhang2001, beniamini2017}. In this scenario the early millisecond spin periods and very strong fields lead to rapid spin-down which can power either GRB-like emission for spin-down times of seconds, or luminous supernovae for spin-down times of weeks \citep{metzger2015, metzger2018}. 

Magnetars are also prime candidates as the origins of Fast Radio Bursts (FRBs). Although it now appears likely that there are at least two classes of such events \citep{pleunis2021}, multiple lines of evidence imply that at least some may be created by magnetars. This includes the strong measures of Faraday rotation in repeating FRBs that imply strong magnetic fields,the location of some well localised FRBs in star forming regions similar to those that host supernovae and GRBs \citep{heintz2020, bochenek2021}, and the galactocentric offset distribution of FRBs being consistent to that of Galactic neutron stars \citep{bhandari2021}. Recently, detailed environmental analysis by \citet{chrimes2021}, when considering the Milky Way as an FRB host, has shown good consistency between distributions of diagnostics for FRBs environments -- such as their location in the light distribution of their host \citep[following][]{fruchter2006} -- and what the equivalent distributions would be for Milky Way magnetars from an extragalactic vantage point.

However, arguably the strongest evidence for an FRB-magnetar link arises from the detection of FRB-like bursts from the Galactic magnetar \sgr \citep{chimefrb2020, bochenek2020, kirsten2021}. These Galactic bursts had similar signatures and durations to extragalactic FRBs, however, when compared the subset of extragalactic FRBs which have been localised to host galaxies, they appear much less luminous \citep[e.g.,][]{nimmo2021}. The luminosity of the \sgr\ bursts means that they would not be detectable in external galaxies, and so the lack of similar luminosity bursts in the extragalactic sample is not surprising. Indeed, the presence of these low-luminosity bursts appears to indicate we are missing the full luminosity distribution of FRBs when accounting only for the extragalactic population. During active episodes, some magnetars emit multiple X-ray bursts (sometimes called burst storms or burst forests) with a very broad range of luminosities. In even rarer cases, magnetars emit Giant Flares (GFs) \citep[e.g.,][]{mazets1979}; only three have been thus far observed in our Galaxy. However, several more have been recently identified to originate from  external galaxies \citep{burns2021}, with properties akin to a short GRB \citep{hurley2005, palmer2005}. Similar behaviour occurring in the radio might yield detectable repeating FRBs from extragalactic magnetars. 

\sgr\ is one of the most active magnetars in the Galaxy and the possibilities it now offers to observe an FRB-emitting system in intricate detail has compounded its astrophysical interest. It is spatially coincident with the supernova remnant \snr\ \citep{gaensler2014}, which is typically inferred to be the result of the magnetar-producing supernova \citep{kothes2018, zhou2020, dosanjos2021}. The supernova itself, based on a study of the remnant, does not appear to be extraordinary by core-collapse standards \citep{zhou2020}, hinting that magnetar production in the deaths of massive stars may be more widespread than initially thought. This scenario is appealing given the current tension for `normal' core-collapse supernovae between the theoretical expectations of purely radioactively-powered explosions \citep{ertl2020, woosley2021} and observations \citep{sollerman2021}.

\sgr\ underwent a new period of increased activity throughout mid-2021, triggering multiple high-energy missions including Fermi \citep{lesage2021GCN30313}, INTEGRAL \citep{mereghetti2021GCN30395}, GECAM \citep{xiao2021GCN30400}, Neil Gehrels \swift\ Observatory \citep{palmer2021GCN30406}, Konus-Wind \citep{ridnaia2021GCN30409} and Calet \citep{nakahira2021GCN30458}, although as yet no FRB-like emission has been reported during this time, nor indeed any detectable radio detection \citep{singh2021}.

Here we present Hubble Space Telescope (\hst) NIR observations of \sgr\ obtained during its recent period of enhanced activity. Our observations confirm the detection of the counterpart with good statistics six years after its previous detection. We discuss the implications of this new detection for the origin of the NIR emission. Additionally, our observations place  constraints  on the proper motion of \sgr, thereby, constraining the kick imparted to the magnetar at birth. 

\subsection{The distance to \texorpdfstring{\sgr}{SGR $1935+2154$}}
\label{sec:distance}

Much of our analysis of \sgr\ imposes a need to know the distance to the object. This distance has been somewhat contentious, usually being measured by proxy of the distance to \snr\ and using several different means, often with discrepant results. Estimates range anywhere from  $\sim$6 to 14\,kpc \citep[e.g.][]{park2013, pavlovic2014, surnis2016, kothes2018, zhong2020a, zhou2020}. For our purposes we will adopt the distance of $6.6\pm0.7$\,kpc determined in a recent study by \citet{zhou2020} of molecular clouds impacted by \snr. The motivation for favouring a lower distance in part comes from a SNR-independent distance estimate to \sgr\ from the detection of an X-ray dust scattering ring by \citet{mereghetti2020}, who estimate a distance of $4.4^{+2.8}_{-1.3}$\,kpc. Even more recently, \citet{bailes2021} estimated a distance to \sgr, based on line-of-sight measures such as column density and extinction, of 1.5--6.5\,kpc -- again consistent with the lower estimates of the SNR. Where appropriate, we also discuss the impact of distance on our results.

\section{Observations and Data reduction}
\label{sec:observations}

New observations presented here were taken with \hst\ WFC3/IR in F140W on June 1 2021 (Programme 16505; PI Levan), covering 4.65\,arcmin$^2$ around \sgr. Total observation time was 2497 seconds, split over 4 equal exposures with sub-pixel dithering. This latest observation will be referred to as epoch \epochfour\ and it follows observations from three earlier epochs of similar observations of \sgr\ which were first presented in \citet{levan2018}, to be referred to as epochs \epochone, \epochtwo, \epochthree.
Since accurate alignment with the previous epochs was desired for the purpose of astrometry, the same observing parameters, including roll angles of the telescope kept in 90$\arcdeg$ steps, were used. In particular, epochs \epochthree\ and \epochfour\ share close to an exact observing repeat, separated by 5 years. Observation details of the epochs are given in Table \ref{tab:observations}.

For each epoch, the four individual \flt\ were drizzle-combined \citep{fruchter2002} using AstroDrizzle within DrizzlePac v3.1.8.\footnote{\url{https://www.stsci.edu/scientific-community/software/drizzlepac.html}} We fixed the final pixel scale to $65$\,mas, i.e.\,roughly halving the native pixel scale. All mentions of pixels throughout are on this pixel scale. The rotation of the drizzled images were set to align with the equatorial coordinates system, such that $\Delta X$ and $\Delta Y$ in image pixel coordinates directly translate to $\Delta\alpha$ and $\Delta\delta$.

\begin{deluxetable}{ccccc}
\label{tab:observations}
\tablecaption{\hst\ WFC3/IR observations of \sgr}
\tablewidth{0pt}
\tablehead{
\colhead{Epoch} &
\colhead{MJD} & 
\colhead{PA$_\textrm{V3}$} &
\colhead{Exptime} &
\colhead{Filter} \\
\colhead{} & 
\colhead{(days)} & 
\colhead{(degrees)} & 
\colhead{(seconds)} & 
\colhead{}
}
\colnumbers
\startdata
\epochone   & 57083.976129  & 115.218903 & 2396.929 & F140W \\
\epochtwo   & 57252.355994  & 295.220612 & 2396.929 & F140W  \\
\epochthree & 57539.962668  &  25.218519 & 2396.929 & F140W  \\
\epochfour  & 59366.721959  &  25.218519 & 2396.929 & F140W  \\
\enddata
\tablecomments{(1) The name of the observation epoch as referred to in the text. (2) Modified Julian Date at midpoint of epoch observation. (3) The position angle of the V3 axis of \hst\ for the first exposure -- this is closely related to the roll angle. (4) Total exposure time of observation. Epochs \epochone, \epochtwo, and \epochthree\ have been previously presented in \citet{levan2018}.
}
\end{deluxetable}

\section{Methods} 
\label{sec:methods}

Throughout we have implicitly assumed the source is point-like in our \hst\ images, any deviation from this would further increase photometric and astrometric uncertainties.

\subsection{Photometry} 
\label{sec:photometry}

Photometry was measured using DOLPHOT V2.0 \citep{dolphin2000}\footnote{\url{http://americano.dolphinsim.com/dolphot/}}, usingg the WFC3/IR package and using updated point spread function (PSF) cores from \citet{anderson2016}. Although drizzled images suffer a number of effects that can compromise their use for photometry, we did use the drizzled image for each epoch as an input reference image for DOLPHOT, as this is used only to initially find sources. For each epoch, the sources are then photometered on the individual (natively-sampled) \flt\ exposures using empirical PSFs, for accurate photometry and astrometry. The final source position and photometric measurements are then determined from a combination of the individual \flt\ measurements. Using the known position of \sgr\ \citep{levan2018} we recover the counterpart in each epoch -- although faint, it is well detected in each epoch, with $\log_{10}(\textrm{Counts}) \sim 3.7$--$4.1$. DOLPHOT natively produces photometry in the VEGAMAG system. To convert to ABMAG, which we will use throughout, we make use of stsynphot\footnote{\url{https://github.com/spacetelescope/stsynphot_refactor}} v1.1.0 to compute the offset between these two systems in the F140W filter, finding the additive correction to be 1.0973\,mag, which we add to each of the output magnitudes from DOLPHOT.

During the course of manual inspection of the frames, it was noticed that the first exposure of epoch \epochtwo\ is affected by what appears to be a cosmic ray. This object is automatically masked by DOLPHOT and means the photometry is significantly compromised for this exposure at the level of offer only a weak detection (SNR $\sim 2$ cf. $\sim 10$ for other exposure in this epoch). Despite the poor constraints, we opt to use this measurement in subsequent analyses, including in the calculation of the final combined photometry for that epoch, but we note in the text wherever this is not the case.

\subsection{Astrometry} 
\label{sec:absolute_astrometry}

With a significant base-line of observations, spanning more than 6 years, we can take advantage of the stability of \hst\ to perform accurate astrometry of \sgr\ in order to place constraints on the proper motion (PM) of the source. As we have no line-of-sight constraints, any discussion of PM ($= \mu$) relates to tangential -- i.e. plane-of-sky -- motion only.

For our astrometric work, we significantly cut the original DOLPHOT photometry tables, per-epoch, for quality. Any source that was not given a ``bright star'' type classification by DOLPHOT was thrown out. This removes some faint stars, as well as any extended objects. Next we set a crowding limit of 0.7\,mag, i.e. we remove any stars for which their magnitude brightens by more than 0.7\,mag if neighbouring stars are not accounted for. This removes a lot of spurious sources and well as preferentially leaving isolated sources. We then cut any object for which DOLPHOT found a zero or negative flux (spurious sources) and finally drop the faintest 10\% of sources that survived previous cuts. This left $\sim$15000 surviving sources per epoch, and a manual inspection of the source catalogues indicated a very high level of commonality of sources between epochs.

The precision of the \gaia\ astrometric solution now offers the opportunity for absolute astrometry to be performed with \hst\ imaging at a comparable level to that found with differential astrometry through tying relative astrometric coordinate systems to the \gaia\ reference frame. The reader is referred to \citet{bedin2018}, hereafter \citetalias{bedin2018}, for a full and thorough pedagogical explanation for the method, which will comparatively only be summarised for our particular use-case here. Where appropriate, we follow their notation described in their section 3.3. (Note that we do not foresee any chance of significant parallax measurements of our source, given its expected distance, and so we do not apply any corrections to compute this value absolutely, cf. \citealt{bedin2020}.)

Firstly, TOPCAT\footnote{\url{http://www.star.bris.ac.uk/~mbt/topcat/}} was used to extract a cone of sources in the \gaia\ Early Data Release 3 \citep[EDR3;][]{gaiaedr3}. From these sources we exclude any that do not have at least a five-parameter astrometric solution computed -- positions and PM in equatorial coordinates as well as a parallax (and pseudo-colour for the six-parameter solutions), and then remove any for which the renomalised unit weight error \citep[RUWE;][]{gaiaedr3_astrom} is greater than 1.2. This cut on RUWE effectively removes those with poor astrometric solutions and is even stricter than that done in \gaia\ EDR3 catalogue validation \citep{gaiaedr3_catval}, as we strongly prioritise quality rather than quantity of the tie points for our astrometry.  
We used the crude World Coordinate System (WCS) information pre-populated in the \hst\ data to convert the position of surviving \gaia\ sources into approximate ($X$, $Y$) pixel coordinates for each epoch, simply for the purpose of doing a very rough and generous first cross-match with a subset of bright sources in the \hst\ images. This provided us with a means to automatically determine a matched list of \gaia\ sources with the accurate ($X$, $Y$) coordinates of their \hst-detected counterparts.
The equatorial \gaia\ coordinates were corrected for their PM in order to obtain their positions as they would have appeared at each of our \hst\ epochs (e.g. see equation 5 of \citetalias{bedin2018}).
Using equation 3 of \citetalias{bedin2018}, epoch-corrected \gaia\ equatorial coordinates were converted to those of a tangent plane with coordinates ($\xi$, $\eta$). This plane is defined with a tangent point of ($\alpha_\circ$, $\delta_\circ$), the values of which were simply chosen to be the reference coordinates of the WCS for each epoch's image. 
These ($\xi$, $\eta$) coordinates of \gaia\ sources could then be further converted into ($X$, $Y$) pixel coordinates for each epoch, following equation 1 of \citetalias{bedin2018}, after an appropriate fitting of the coefficients.

Using the matched \gaia\ and \hst\ source lists, we initially fitted for the linear coefficients of equation 1 in \citetalias{bedin2018} using the \gaia\ ($\xi$, $\eta$) coordinates, and the \hst\ ($X$, $Y$) coordinates in a weighted least sqaures (Levenberg-Marquardt) manner using lmfit \citep{lmfit}. From this initial fit we identified any matches with an offset larger than \percentile{95} of the distribution to prevent outliers affecting the fitting, and removed those before repeating the fit to obtain our final fitted parameters. Practially, the exact choice of percentile rejection made little difference to results, so long as the few most discrepant sources, which were noticeable poorer than the overall distribution of offsets for some epochs, were removed for each fit. The nominal value of parameters and uncertainties from the final fits were determined by propagating \gaia\ astrometric uncertainties in position and PM onto the ($\xi$, $\eta$) coordinates, and repeating the fitting 2000 times with re-sampled realisations of these coordinates. The median and standard deviation of these results gave our results. For the linear parameters we also add in quadrature the median statistical uncertainty on the fit into our uncertainty budget for these values. 

For the \gaia--\hst\ removed matches (i.e. those with offsets $>\percentile{95}$ after the initial fit), manual inspection did not reveal any strongly obvious reason for their larger offsets in terms of crowding, chip location etc. The \gaia\ astrometric parameters from these sources were typically larger than the overall source matches, and could be indicative of incorrect and lack-of treatment of the PM and parallax, respectively.

It is worth noting at this point that we would ideally wish to use solely Quasi-Stellar Objects (QSOs) for our tie points -- point-like sources with practically zero PM and parallax. In this case the differential movement of our tie points between epochs in the \gaia\ reference frame would not be a concern. However, owing to sky density of QSOs, and the Galactic location of \sgr\ in the plane, we have no such objects within our field of view.

\section{Results}
\label{sec:results}

\subsection{Photometry}
\label{sec:photometry_results}

\begin{figure*}
\plotone{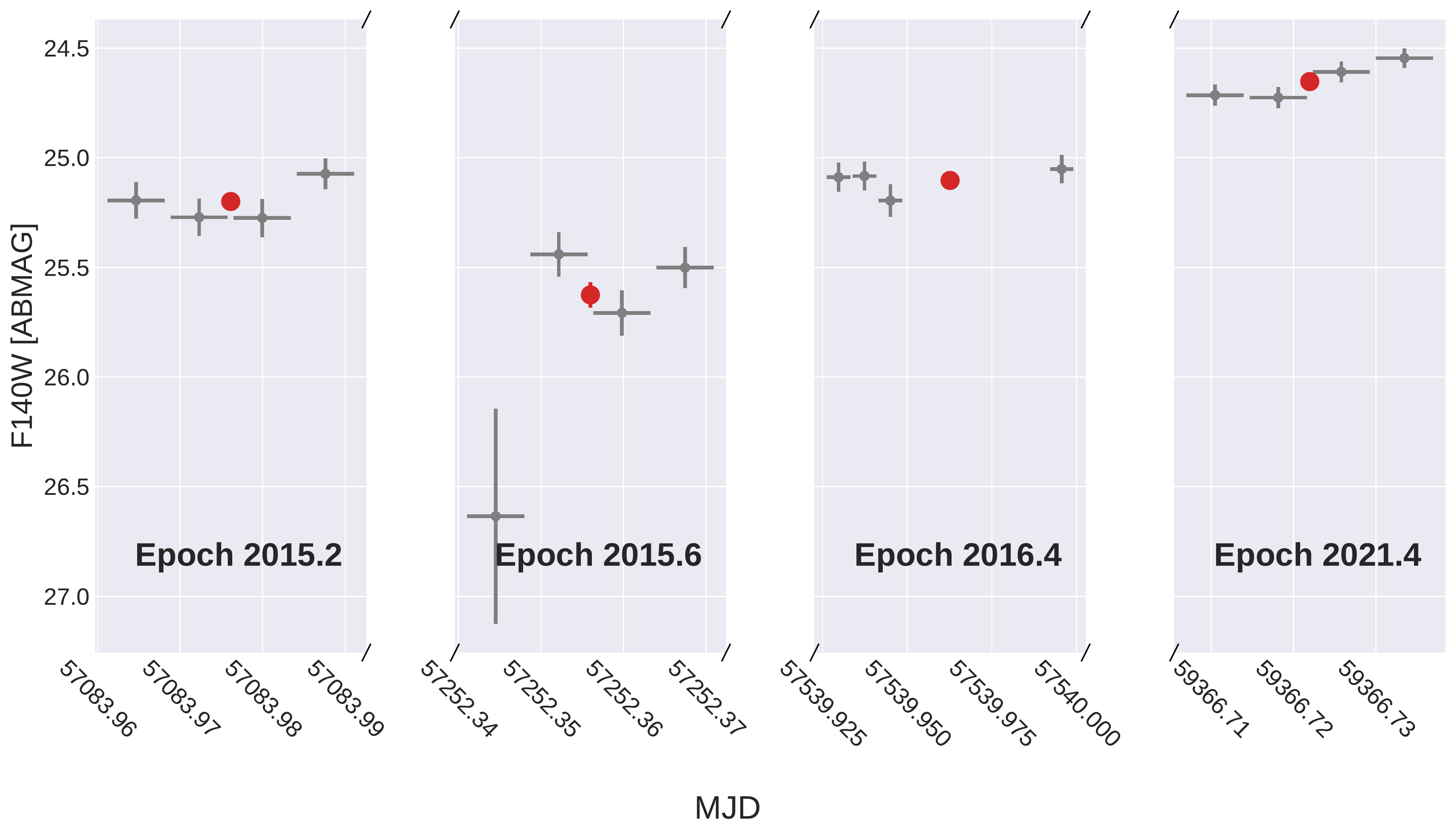}
\caption{
\label{fig:sgr_light_curve}
The \hst\ WFC3/IR F140W light curve of \sgr. Small grey points indicate photometry from individual \flt\ data, larger, blue points indicate the combined photometry per epoch. The variability of \sgr\ on long timescales (note the broken abscissa axis) is apparent from the deviation of the blue points, although intra-epoch variability is less clear (see text). As discussed in the text, the errant measurement in the first exposure of epoch \epochtwo\ was caused by a cosmic ray hit on top of the \sgr\ in the data.}
\end{figure*}

Photometry of \sgr, calculated as detailed in Section \ref{sec:photometry}, is presented in Table \ref{tab:photometry} and shown in Figure \ref{fig:sgr_light_curve}. \sgr\ is significantly variable on timescales of years and has undergone a factor $\sim2.5$ increase in brightness between Epochs \epochtwo\ and \epochfour. In addition to this, we investigated the possibility of variability on short (minutes) timescales. For this we determined $p$-value for the null hypothesis of no variability by modelling the magnitudes as a constant. The first three epochs, \epochone, \epochtwo, and \epochthree, provide $<2\sigma$ rejection of this null hypothesis (we trialled \epochtwo\ with and without the compromised first exposure), with \epochfour\ rejecting at $p=0.016$. Overall we find no strong evidence for short timescale variability of \sgr\ in current observations, with future observations -- especially if the IR counterpart remains brighter, which allows for more precise photometry -- needed to rule more conclusively. 

\begin{deluxetable}{ccccc}
\label{tab:photometry}
\tablecaption{\hst\ WFC3/IR photometry of \sgr}
\tablewidth{0pt}
\tablehead{
\colhead{Epoch} &
\colhead{MJD} & 
\colhead{Data set} &
\colhead{\magf} & 
\colhead{$\sigma$(\magf)} \\
\colhead{} & 
\colhead{(days)} & 
\colhead{} & 
\colhead{(AB mag)} & 
\colhead{(AB mag)} \\
}
\colnumbers
\startdata
             & 57083.964654  & icst01htq  & 25.194 & 0.083 \\
             & 57083.972304  & icst01huq  & 25.271 & 0.085 \\
\epochone    & 57083.979955  & icst01hwq  & 25.274 & 0.087 \\
             & 57083.987605  & icst01hxq  & 25.073 & 0.071 \\
             & 57083.976129  & combined   & 25.199 & 0.041 \\
\hline
             & 57252.344519  & icst01htq  & 26.634 & 0.491 \\
             & 57252.352169  & icst01huq  & 25.440 & 0.101 \\
\epochtwo    & 57252.359820  & icst01hwq  & 25.707 & 0.104 \\
             & 57252.367470  & icst01hxq  & 25.501 & 0.094 \\
             & 57252.355994  & combined   & 25.625 & 0.058 \\
\hline
             & 57539.929757  & icst01htq  & 25.088 & 0.066 \\
             & 57539.937408  & icst01huq  & 25.083 & 0.066 \\
\epochthree  & 57539.945058  & icst01hwq  & 25.195 & 0.075 \\
             & 57539.995579  & icst01hxq  & 25.051 & 0.065 \\
             & 57539.962668  & combined   & 25.103 & 0.034 \\
\hline
             & 59366.710483  & icst01htq  & 24.714 & 0.048 \\
             & 59366.718134  & icst01huq  & 24.725 & 0.048 \\
\epochfour   & 59366.725784  & icst01hwq  & 24.608 & 0.048 \\
             & 59366.733435  & icst01hxq  & 24.545 & 0.045 \\
             & 59366.721959  & combined   & 24.652 & 0.024 \\
\enddata
\tablecomments{(1) The name of the observation epoch as referred to in the text. (2) Modified Julian Date at midpoint of exposure.  (3) The \flt\ data set name of the exposure as retrieved from the \hst\ archive -- `combined' indicates the combined photometry for each epoch. (4) and (5) The AB magnitude and uncertainty of \sgr.}
\end{deluxetable}

\subsection{Astrometry} 
\label{sec:absolute_astrometry_results}

Our final fitted parameters from the absolute astrometry, and associated fit statistics, are given in Table \ref{tab:absolute_astrometry_coeffs}. The $\sigma(\Delta X)$ and $\sigma(\Delta Y)$ values, which give the typical spread of offsets between \hst\ image source positions and transformed \gaia\ equatorial coordinates in the ($X$, $Y$) pixel coordinate system, are comparable or lower than that obtained from differential astrometry of \hst\ frames.\footnote{See, for example, \citet{levan2018}. Our own differential astrometric exercises with these data using spalipy \citep{spalipy} for astrometric alignment provided comparable, but slightly larger, alignment residuals -- this prompted us to concentrate solely on the absolute astrometry.} Firstly, we can conclude that the precision of the \gaia\ astrometric solution is not a limiting factor in our transformation, and that the move to an absolute reference frame is possible without sacrificing precision cf. differential astrometry. We further conclude that strict selection of tie-points for absolute astrometry, as detailed in Section \ref{sec:absolute_astrometry}, and the ability to correct for the known PM of tie point sources, are significant aids in the accuracy of the fit results.

\begin{figure}
\includegraphics[width=\linewidth]{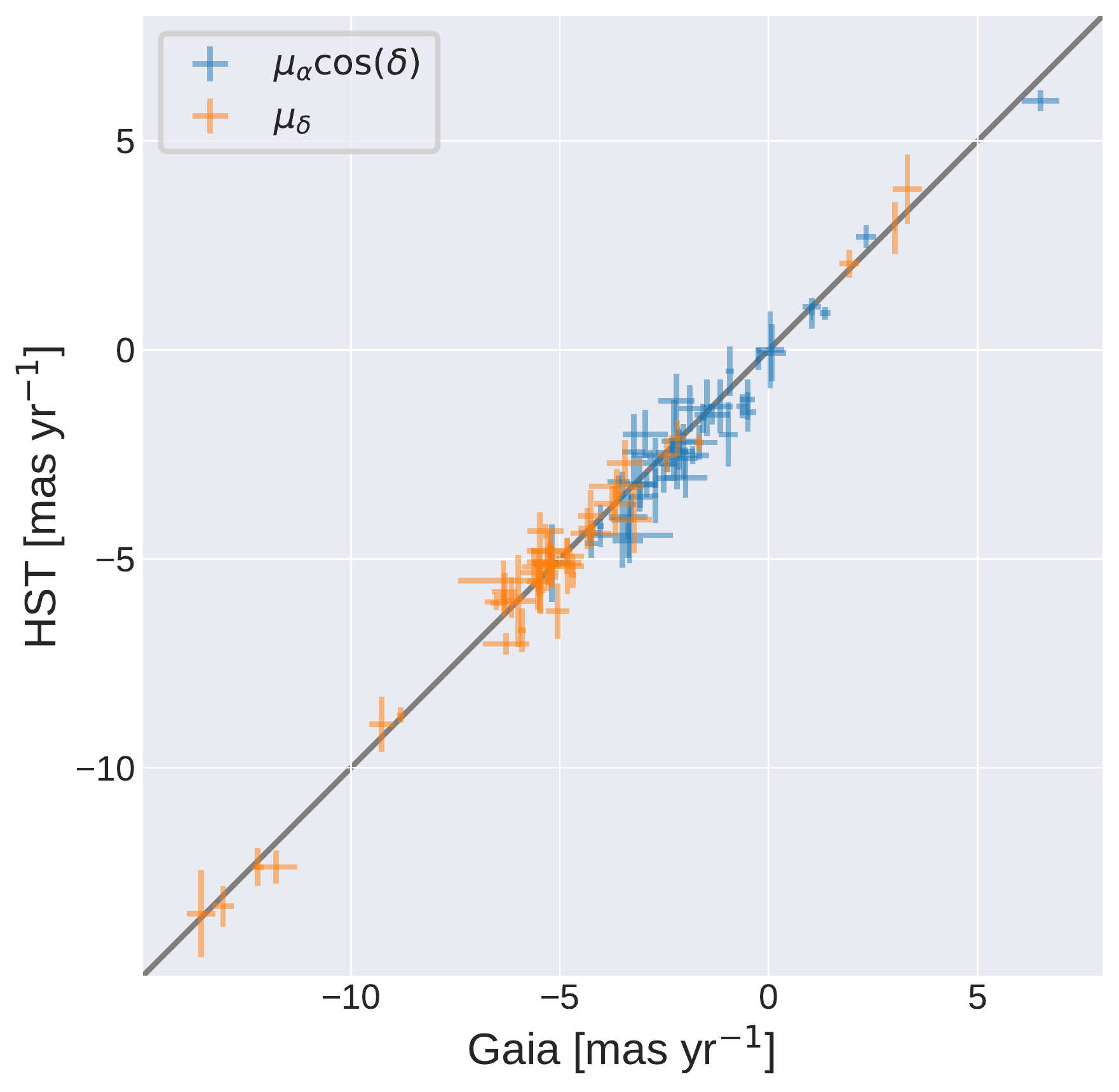}
\caption{
\label{fig:pm_gaia_hst_compare}
The PM constraints for 50 sources in the centre of the \hst\ field of view from our own \hst\ calculations, and those of \gaia\ EDR3. We observe excellent agreement, with no evidence of a systematic shift in either coordinate at high or low absolute values.
}
\end{figure}

As a sanity check for our transformations and subsequent PM calculations, we compared results from our routine to calculate PMs from \hst\ data with \gaia\ EDR3 PMs for the 50 brightest sources in a region at the centre of the \hst\ images that is covered by all 4 epochs. The results of this are shown in Figure \ref{fig:pm_gaia_hst_compare}, which highlights excellent agreement.

\begin{figure*}
\includegraphics[width=\linewidth]{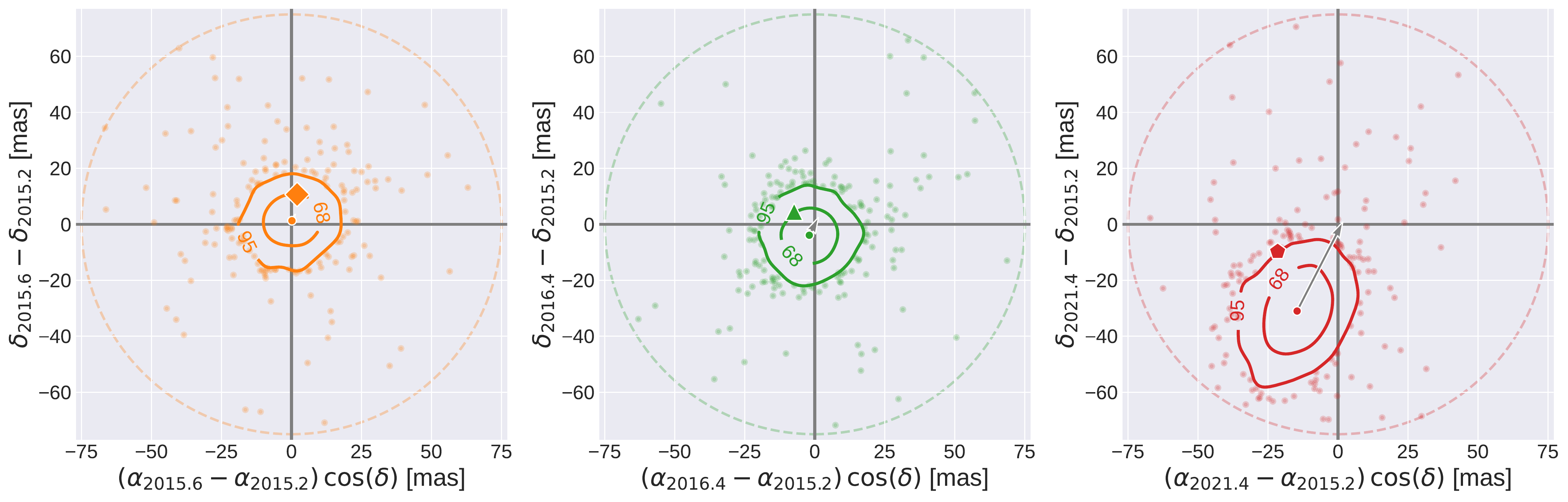}
\caption{
\label{fig:absolute_pm_contours_close_mag}
Equatorial offsets of sources between different \hst\ epochs in the absolute astrometric reference frame of \gaia. Equatorial coordinates on the axes labels are denoted with a subscript indicated the epoch they refer to. In each subplot the position of \sgr\ is given by the large coloured marker.
Overlaid are contours of 68 and 95 percentiles (\percentile{x}; determined via a Gaussian kernel-density estimator), showing the distribution of offset values for all sources in the image that are close in brightness to \sgr. Those sources outside \percentile{95} are individually plotted. The peak of the distribution is marked by a small circle marker, and grey lines guide the eye to the location of zero offset. A grey vector is added to each subplot (although not visible on the left subplot), which denotes the direction and magnitude of motion expected from a model of differential Galactic rotation in the direction of \sgr\, for a source 8.5\,kpc away -- the distance of the Perseus arm of the Galaxy along this sight-line.
The large dashed ellipses mark the offset limit, based on the tolerance of source matching used (see Section \ref{sec:absolute_astrometry_results}). 
}
\end{figure*}

We used our absolute astrometry to perform a cross-match of all sources similar in brightness ($23.5 < \magf < 24.5$\,mag) to \sgr\ -- around 2700 sources per epoch. For this cross-match we converted all ($X$, $Y$) coordinates of the sources into the \gaia\ Equatorial frame in ($\alpha$, $\delta$) using our fitted parameters, and then matched our sources in Equatorial coordinates between epochs, allowing a generous matching radius of 75\,mas (the reason for which will become apparent). The results of these cross matches are shown in three sub-plots of Figure \ref{fig:absolute_pm_contours_close_mag}. What is immediately clear is the systematic shift in the offsets as the baseline of time between epochs increases. This indicates bulk motion of the field, and to determine the \emph{peculiar} velocity of \sgr\ within this bulk motion, it must be accounted for. We surmised this to be a manifestation of differential Galactic rotation and thus sought to model its effect.

Sources at different sight-lines and distances throughout the Galaxy will have a Galactic PM associated with them, which arises as a combination of their rotation in the Galaxy and the solar peculiar motion, each with respect to the local standard of rest (LSR). We follow largely the procedure in \citet[][section 3.2]{verbunt2017}, albeit with updated parameters. The velocity of the Sun with respect to the LSR is taken to be ($U$, $V$, $W$) = ($8.63\pm0.64$, $4.76\pm0.49$, $7.26\pm0.36$)\kms\ \citep{ding2019}, with the components of motion being towards the Galactic centre, along Galactic rotation, and perpendicular to the Galactic plane, respectively. We use a constant rotation velocity, $v_R = 220$\kms\ (IAU standard) for both the LSR and our field, since all sight-line distances have Galactocentric distances outside the turnover to a flat rotation speed profile for the Milky Way \citep[$\sim$\,3kpc, e.g.][]{reid2014}. Finally we set the Sun's Galactocentric distance as $8.122\pm0.033$\,kpc \citep{gravity2018}. Models such as this have also been employed elsewhere in the pursuit of PM constraints for other similar objects \citep[e.g.][]{dodson2003, deller2012, tendulkar2012}.

The sight-line to \sgr\ is dominated by the Perseus arm of the Milky Way, at a distance of $\sim8$\,kpc \citep{vallee2008} and in Figure \ref{fig:absolute_pm_contours_close_mag} we show vectors of bulk motion of the field due to differential Galactic rotation at this distance. The vectors, particularly for our longest base-line between epochs in the third sub-plot, quite accurately account for the shift in the peak of the distributions from the origin in these plots. With the model verified, we are therefore able to quantify the effect of Galactic motion on \sgr\ (for a given distance) and therefore, by removing this Galactic motion, we can derive its peculiar velocity -- the quantity of interest to study the nature of its natal kick.

\begin{figure*}
\includegraphics[width=\linewidth]{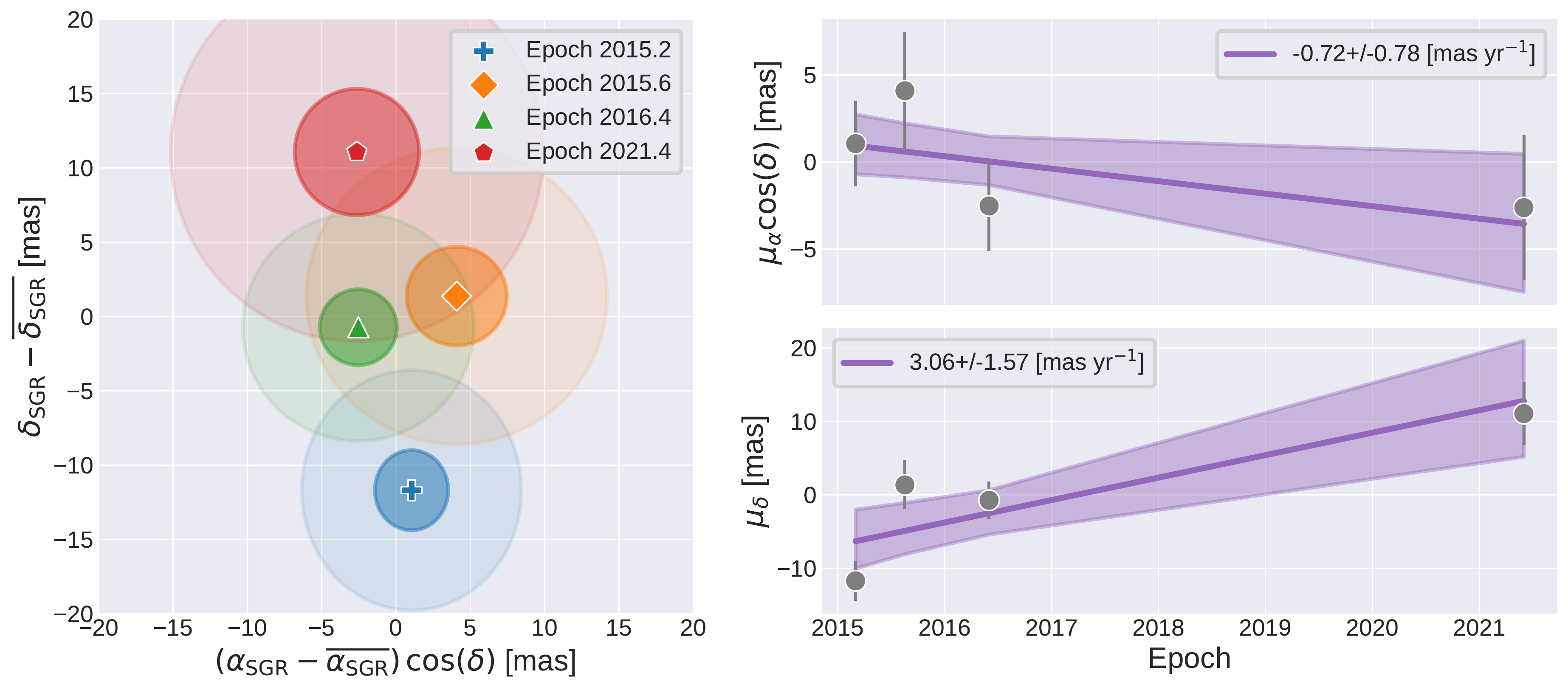}
\caption{
\label{fig:absolute_pm_slope_6.6}
{\em Left:} Equatorial offsets of \sgr\ in the \gaia\ absolute frame for our \hst\ epochs, relative to its mean position. These offsets have been corrected for Galactic motion of \sgr\ due to differential Galactic rotation and the solar peculiar velocity, assuming a distance of $6.6\pm0.7$\,kpc to the SGR. One and three sigma uncertainties on the positions are given by the darker and lighter ellipses surrounding each point. {\em Right:} Linear fitting to equatorial offsets of \sgr. The uncertainties of each offset (grey markers) include contributions from the positional uncertainty on \sgr, the offset residuals of the astrometric tie to the \gaia\ frame (Table \ref{tab:absolute_astrometry_coeffs}), and the uncertainty on the Galactic motion correction. The value of fitted slope, which gives the PM constraint, is shown by the legend in each sub-figure, with the shaded area indicating the $1\sigma$ uncertainty on the model.
}
\end{figure*}

Following Section \ref{sec:distance}, we use $6.6\pm0.7$\,kpc as our adopted distance to \sgr. Accounting for uncertainties both in the model parameters (where given), as well as the distance, we derive a Galactic PM for \sgr\ of $(\pmrag, \pmdecg) = (-2.64\pm0.04, -5.20\pm0.04)$\masyr. The remaining motion of \sgr\ in our absolute frame is its peculiar motion with respect to its own local standard of rest. In Figure \ref{fig:absolute_pm_slope_6.6} we show the position of \sgr\ in each epoch relative to the mean position across all four epochs: $(\overline{\alpha_\textrm{\tiny SGR}}, \overline{\delta_\textrm{\tiny SGR}}) = (293.73170336, 21.89657038)$\,deg. A linear model was constructed and fitted in a weighted least-squares manner to the relative motion of \sgr. For the offsets' uncertainties, the positional uncertainty on the \sgr\ itself, the typical offset residuals after alignment to the \gaia\ frame (Table \ref{tab:absolute_astrometry_coeffs}), and the uncertainty on the Galactic PM were all added in quadrature as sources of random statistical error. Out fitting gave $(\pmra, \pmdec) = (-0.73\pm0.74, 3.03\pm1.55)$\masyr, i.e. a total PM of $\pmtot = 3.12\pm1.52$\masyr, or motion at the $\sim 2.1\sigma$ level of confidence. We note that our fitting procedure would ideally be performed on fully homoscedastic data consisting of a large number of datapoints. Unfortunately, we are limited to only four data points for our fitting (although noting each point are themselves effectively averages constructed from four independent measurements in the individual \flt\ files), and as such the central-limit theorem is not in action -- this makes translating the $\sim 2.1\sigma$ detection of motion to a formal probability more difficult. With these caveats in mind we work with the above value for the PM of \sgr\ as the current best available constraints, and finally note even if we were to consider this a non-detection, given our low uncertainties, our discussions and conclusions remain unchanged. Figure \ref{fig:pm_gaia_hst_compare} gives further credence that the fitting procedure produces results in line with those determined from a richer astrometric dataset.

\begin{figure}
\includegraphics[width=\linewidth]{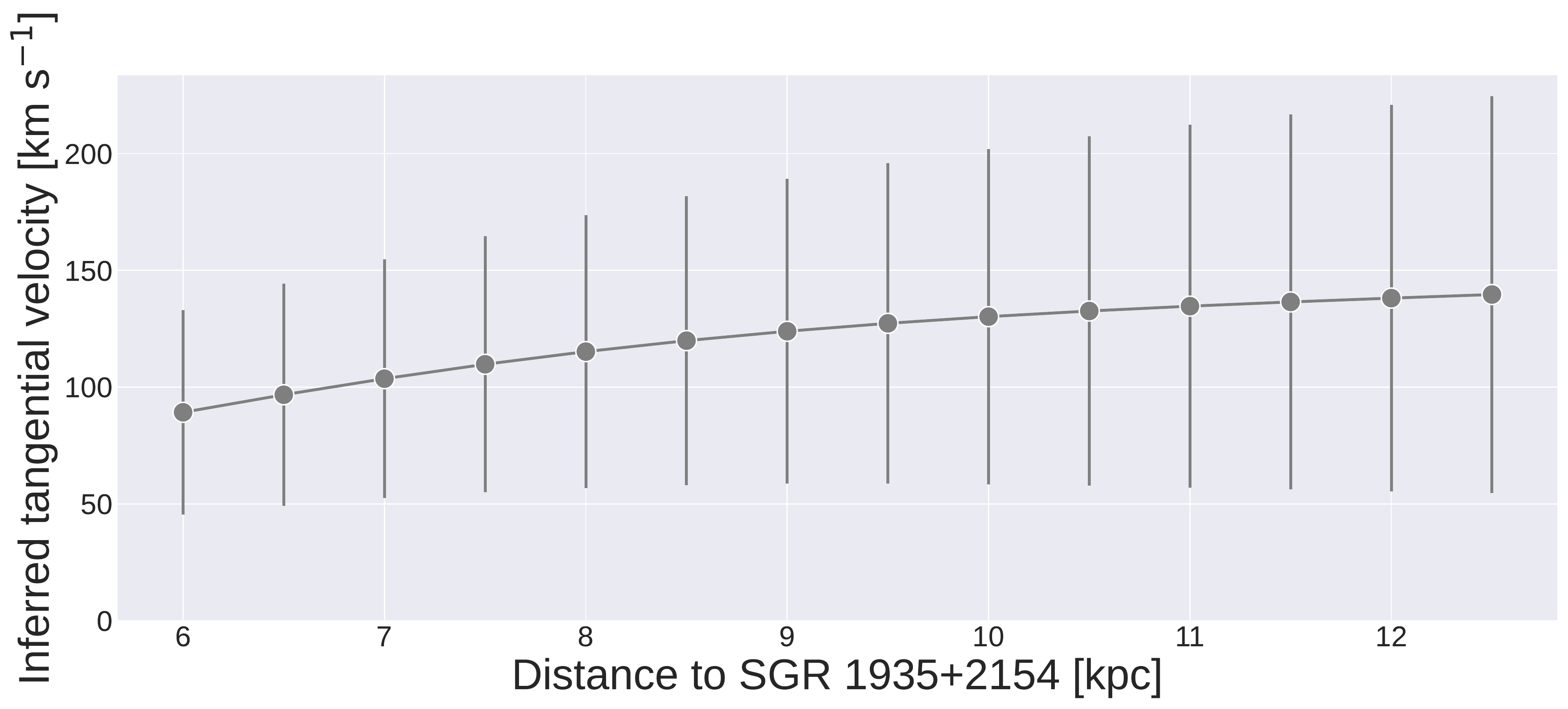}
\caption{
\label{fig:velocity_vs_distance}
The inferred tangential space velocity of \sgr\ for a range of assumed values for the distance. The non-linear variation is a result of the differential Galactic rotation model, which is used to subtract of the bulk field motion before calculating the peculiar motion of \sgr\ (Sections \ref{sec:absolute_astrometry} and \ref{sec:absolute_astrometry_results}).
}
\end{figure}

Since our corrections for Galactic motion are distance-dependent the tangential space velocity in units of \kms is not a simple function of assumed distance to \sgr. In Figure \ref{fig:velocity_vs_distance} we show the results for a variety of assumed distances to \sgr, and note that the overall change based on assumed distance is largely dwarfed by the astrometric uncertainty. Taking two example distance estimates at the lower and upper bounds from the literature -- $6.6\pm0.7$\,kpc \citep{zhou2020} and $12.5\pm1.5$\,kpc \citep{kothes2018} -- we obtain tangential space velocities of $\vtot\sim97\pm48$ and $140\pm89$,\kms, respectively (note these values include the additional uncertainty contribution from the distance assumed).\footnote{For the distance estimate of \citet{kothes2018}, other corresponding values are: $(\pmrag, \pmdecg) = (-2.24\pm0.14, -4.22\pm0.28)$\masyr, $(\pmra, \pmdec) = (-1.11\pm0.75, 2.09\pm1.60)$\masyr, and $\pmtot = 2.36\pm1.48$\masyr.}

\begin{deluxetable*}{ccccc}
\label{tab:absolute_astrometry_coeffs}
\tablecaption{Fitted linear coefficients and fitting statistics for transformations between the ($X$, $Y$) pixel coordinate system of each epoch and a tangent plane ($\xi$, $\eta$) defined in the \gaia\ EDR3 reference frame.
}
\tablewidth{0pt}
\tablehead{
\colhead{Value} & 
\multicolumn{4}{c}{Epoch} \\
\cline{2-5}
\colhead{} &
\colhead{\epochone} &
\colhead{\epochtwo} &
\colhead{\epochthree} &
\colhead{\epochfour}
}
\colnumbers
\startdata
\hline
\multicolumn{5}{c}{Linear parameters} \\
\hline
$\mathcal{A}$ [mas]        & $-65.0081\pm0.0003$  & $-65.0024\pm0.0004$  & $-65.0077\pm0.0003$  & $-65.0039\pm0.0011$ \\
$\mathcal{B}$ [mas]        & $0.0037\pm0.0003$    & $0.0010\pm0.0005$    & $0.0019\pm0.0003$    & $0.0036\pm0.0010$    \\
$\mathcal{C}$ [mas]        & $0.0040\pm0.0003$    & $0.0004\pm0.0004$    & $0.0018\pm0.0003$    & $0.0032\pm0.0011$    \\
$\mathcal{D}$ [mas]        & $65.0058\pm0.0004$   & $65.0012\pm0.0005$   & $65.0092\pm0.0003$   & $65.0038\pm0.0011$   \\
$\mathcal{X}_\circ$ [pix]  & $1249.473\pm0.003$   & $1492.008\pm0.004$   & $1260.481\pm0.003$   & $1257.427\pm0.009$  \\
$\mathcal{Y}_\circ$ [pix]  & $1092.708\pm0.003$   & $1174.197\pm0.004$   & $1331.946\pm0.003$   & $1331.024\pm0.009$  \\
$\alpha_\circ$ [deg]       & 293.72961531         & 293.73437466         & 293.7312987          & 293.7312413 \\
$\delta_\circ$ [deg]       & 21.89747904          & 21.89599149          & 21.8944872           & 21.8945040  \\
\hline
\multicolumn{5}{c}{Fitting statistics} \\
\hline
Initial \gaia--\hst\ matches           & 128              & 125              & 118              & 117  \\
Clipped matches                       & 7                & 7                & 6                & 6  \\
$\sigma(\Delta X)$ [pix]              & $0.025\pm0.001$  & $0.043\pm0.001$  & $0.027\pm0.001$  & $0.075\pm0.006$ \\
$\sigma(\Delta Y)$ [pix]              & $0.031\pm0.001$  & $0.043\pm0.001$  & $0.027\pm0.001$  & $0.081\pm0.008$ \\
Offset residual \percentile{68} [mas] & $2.6$            & $4.1$            & $2.8$            & $5.2$ \\
Offset residual \percentile{99} [mas] & $4.8$            & $7.5$            & $4.2$            & $10.7$ \\
\enddata
\tablecomments{(1) The name of the value. Linear parameter definitions follow those of \citetalias{bedin2018}; fitting statistics denote, respectively: The number of initial matches used to fit the linear transformation, the number of clipped matches before doing the final fit (i.e. those with offsets $>\percentile{95}$ after the initial fit), the standard deviation of the $X$ and $Y$ pixel offsets of matched sources, and their 68 and 99 percentile total offset residuals in mas. (2--5) The value in the specific epoch's fit.
}
\end{deluxetable*}

\section{Discussion}
\label{sec:discussion}

\subsection{Photometric evolution and origin of NIR emission}
\label{sec:niremission}

\begin{figure*}
\includegraphics[width=\linewidth]{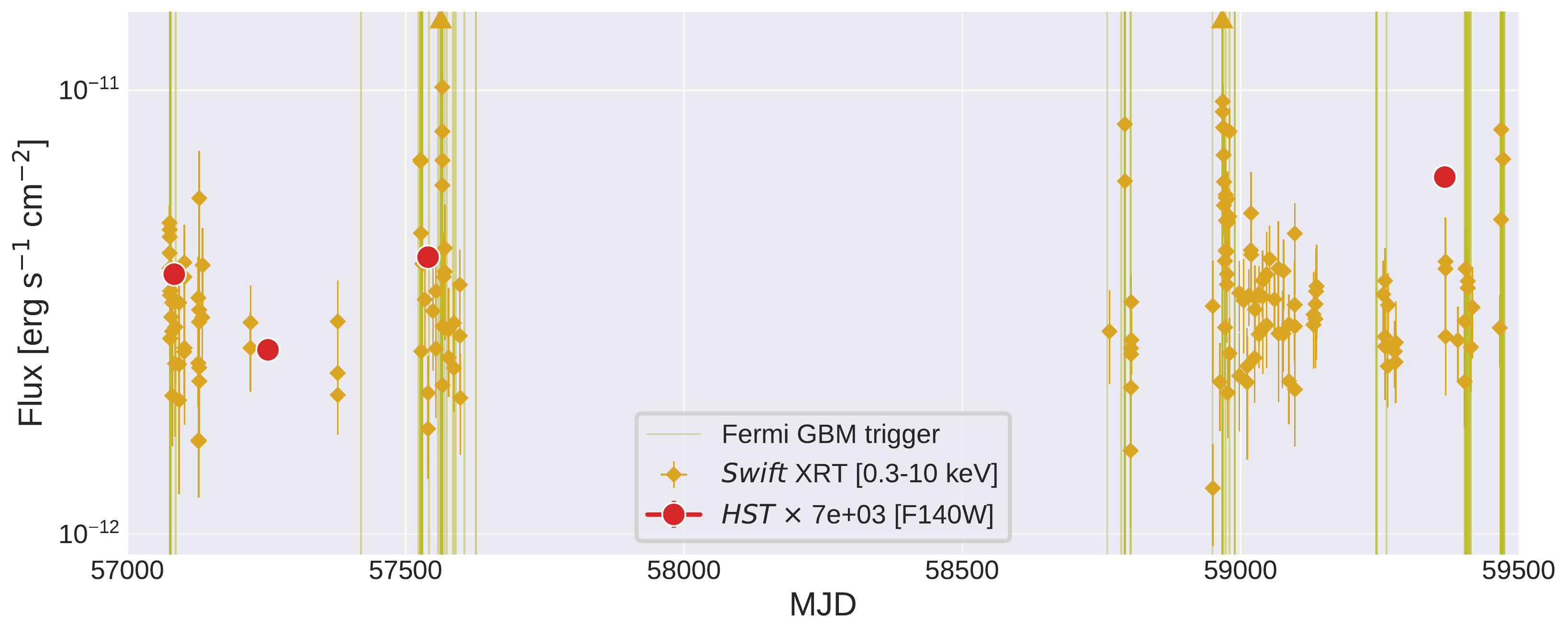}
\caption{
\label{fig:high_energy_lc}
The light curve of \sgr\ in the NIR (\hst\ F140W) and X-ray (\swift\ XRT). Overlaid at the times for Fermi GBM triggers in the vicinity. The latest \hst\ epoch was taken during a period of relative quiescence, in terms of high energy activity, yet is significantly ($\sim$0.5 to 1\,mag) brighter than previous epochs. The intensive mid--late 2021 activity appears to the right of the plot in the form of a high density of Fermi GBM triggers and renewed X-ray activity. (Note for clarity, some high flux \swift\ XRT points have been clipped, and are indicated by upward triangles at the top of the plot.)}
\end{figure*}

The most recent epoch of \hst\ imaging at \epochfour\ shows a factor 1.5 to 2.5 brightness increase in \sgr\ over earlier epochs (\epochone--\epochthree) as shown in Figure \ref{fig:sgr_light_curve}. This latest observation was timed close to a period of increased high-energy activity from the source. We show in Figure \ref{fig:high_energy_lc} our \hst\ photometry alongside the X-ray light curve from the Neil Gehrels \swift\ Observatory X-ray Telescope \citep[XRT; built using the tools described in][]{evans2007,evans2009}, and Fermi Gamma-ray Burst Monitor \citep[GBM;][]{fermigbm2009} trigger times\footnote{Fetched from \url{https://heasarc.gsfc.nasa.gov/W3Browse/fermi/fermigtrig.html}.} in the vicinity of \sgr. 
The \hst\ observation for Epoch \epochfour\ occured on 1 June 2021, with one of the first bursts from the period of mid-2021 activity being reported by Fermi on 24 June \citep{lesage2021GCN30313}. As can be seen from contemporaneously scheduled \swift\ XRT observations, at Epoch \epochfour\ the source appears to be at its quiescence level, and there had been no prior triggers. This perhaps indicates the NIR brightness changes are not correlated to changes at higher energies. Such behaviour in the most recent epoch is at odds with earlier findings based on the first three epochs \citep{levan2018}, also shown in Figure \ref{fig:high_energy_lc}, where NIR brightness-state appears more closely linked to the high-energy activity. Any disconnect may indicate separate emission mechanisms for the two energy regimes, but a statistically rigorous analysis of the cohesion of multi-wavelength activity requires a significant number of additional epochs of NIR monitoring -- both during active and quiescent periods.

Our searches found no significant signs of variability within each epoch's observation. A $p=1.6\%$ rejection of the null-hypothesis of no variability for the latest, and brightest, epoch does perhaps offering some marginal indication that warrants further investigation. Although short period variability in SGRs on the timescales of their magnetars' rotation periods (seconds) have been found \citep[e.g.,][]{kern2002, dhillon2011}, the origin of any minutes--hour long time-scale variability would be less obvious. Additional observations while the source is brighter (which enable more precise photometry), have a better chance to rule on the presence of variability of the source over these timescales. 

One explanation for the origin of the NIR counterpart at the location for magnetars is emission from a debris disk \citep[e.g.][]{perna2000}, heated by X-ray emission from the magnetar itself. Such a disk is claimed to power the emission seen in 4U\,0142 \citep{wang2006}. The NIR emission associated with 4U\,0142 shows variability at the $\sim0.5$ magnitude level, comparable to \sgr. However, in the disk scenario, the NIR is expected to vary closely in sync with changes in X-ray activity \citep[e.g.,][]{rea2004}. This is not obviously the case with \sgr\ (Figure \ref{fig:high_energy_lc}), particularly given the most recent \hst\ epoch where the NIR emission is at its brightest during an X-ray quiescent period. Such a lack of correlation between frequency bands has been seen in magnetars with much richer NIR and X-ray data \citep[e.g.,][]{durant2006}.

Although direct thermal surface emission from the magnetars cannot account for optical/NIR emission due to the unrealistic brightness temperatures inferred, an origin for the emission in the magnetosphere is a possibility. In this scenario, shearing of the magnetic field, caused by starquakes, populates a hot plasma corona surrounding the magnetar \citep[][and references therein]{beloborodov2007}. Although models for the NIR emission are able to reproduce the observed levels of emission seen in magnetars \citep{zane2011}, relatively little is discussed about the time-scales or amplitudes of any variability. Nonetheless, for a magnetosphere origin, emission would be expected to vary roughly concurrently across bands \citep{tam2008}.

Given the prevalence of binaries systems among massive stars \citep[e.g.][]{sana2012}, it is likely \sgr\ did (or does) reside in a binary. The prospects of a binary companion as the NIR emission source require the binary survived the mass-loss and natal kick brought about by the supernova. However, stellar companions would be expected to be much brighter than our detection: for a roughly equal mass binary, assuming an O9V star spectrum \citep{pickles1998} with $M_V = -4.49$\,mag undergoing $A_V = 6.6--8.8$ \citep{green2019}\footnote{The spread in extinction values is largely governed by whether a dust cloud at a similar distance to our adopted distance of \sgr\ is included.} at a distance of $6.6\pm0.7$\,kpc, we might expect it to appear as a $\magf \sim 13$ to $13.5$\,mag source, far brighter than any current detection. A extremely low mass companion -- and consequently extreme initial mass ratio binary -- would need to be inferred to remain compatible with the brightness of the NIR counterpart. For this reason we disfavour a companion as the origin of the emission. A fuller discussion of magnetar binary companions, including \sgr, is presented by Chrimes et al. (in prep).

\subsection{Implications for progenitor from proper motion}

Using two methods to measure the tangential velocity of \sgr, we obtained no strongly significant indications of motion, with a $2.1\sigma$ limit on this velocity of $\lesssim100$\kms. The observed velocity distribution of NSs, measured primarily from pulsars, is quite poorly understood, particularly so for the presence and relative contribution of any low velocity component \citet[e.g.][]{arzoumanian2002, brisken2003, hobbs2005, verbunt2017}. However, even \citet{hobbs2005}, where the overall velocities are fit with a single, wide Maxwellian distribution and no low-velocity component, we find that $\sim20\%$ of the transverse velocities of young (characteristic age $<3$\,Myr) are $\leq100$\kms. Our PM constraints therefor place \sgr\ as relatively low velocity, at least compared to the overall pulsar distribution, but not exceptionally so. It is also equal to the lowest tangential velocities found for the few magnetars with such constraints. Further comparison of magnetar and pulsar kick distributions is given in Section \ref{sec:compare_kick_distribution}.

There are astrophysical reasons one may expect a measure of bi-modality in the velocity distribution of NSs. Firstly, the comparative prevalence of low mass stars\footnote{Here `low' is a relative term, used within the domain of stars that are able to produce a NS at their end of their lives in a supernova.} and the relatively sharp lower mass limit cut-off at $M_\textrm{\tiny ZAMS} \simeq 8$\msun for a supernova \citep[e.g.][]{smartt2009}, should result in a ``pile-up'' of low ejecta mass events, in which the imparted SN-kick on the NSs is correspondingly small \citep[given some relation between SN ejecta mass and imparted kick; e.g.][]{bray2016}. The less-numerous tail of higher mass SN progenitors, with correspondingly larger ejecta masses would then contribute to a wider, high-velocity component. Secondly, binary interactions will contribute to this bi-modality. If the NS is borne of a SN in a pre-existing binary, the velocity will be lower than if the SN progenitor's binary had already been disrupted due to the previous SN of its companion -- in these cases the SN progenitors may already have a significant velocities, increasing the mean and spread of any resultant NS velocity distribution produced from such stars. The low velocity of \sgr\ would then hint towards a lower-mass progenitor and/or it being the primary supernova in any putative binary system to which it belongs (with implications for companion emission searches -- Section \ref{sec:niremission}). We must caveat such discussion with our lack of line-of-sight information, meaning we have no constraints on the component of the velocity along this axis.

There are alternative, non-core-collapse, theoretical models to create magnetars -- these typically involve mergers or accretion in systems including white dwarfs and/or neutron stars \citep{margalit2019, zhong2020b}. Given the association to \snr\ (see Section \ref{sec:snr_association}), we have above concentrated on implications given a core-collapse origin.

\subsection{Origin and association to \texorpdfstring{\snr}{SNR G57.2+08}}
\label{sec:snr_association}

The typically young span inferred for magnetar life-times arises from their associations with SNRs, which can be age-dated \citep{allen2004}, and their links to nearby clusters, which, given PM constraints for the magnetar, can provide age constraints by tracing the PM back in time to the cluster \citep[e.g.][]{tendulkar2012}. Such ``kinematic'' ages are typically more reliably than ``characteristic'' ages derived from the spin-down of the magnetic field, which include a number of overly simplifying assumptions about the magnetic field and its evolution \citep{vigano2013}. Kinematic ages for magnetars are typically $10^3$ to 10$^4$\,yr.

\begin{figure*}
\includegraphics[width=\linewidth]{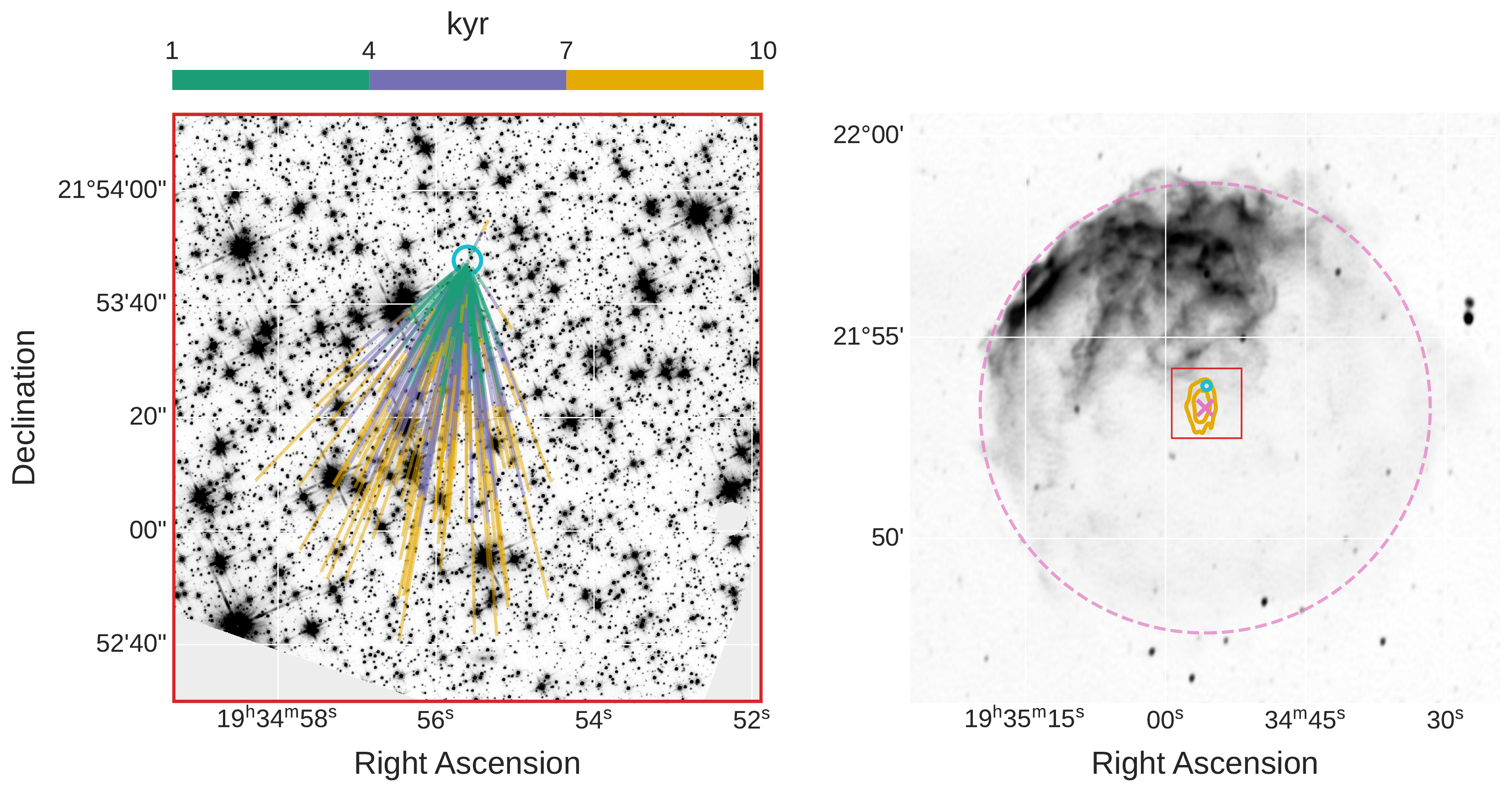}
\caption{
\label{fig:sgr_origin_snr}
{\em Left:} The position and origin of \sgr\ in our epoch \epochfour\ \hst\ WFC3/IR image. The observed location of \sgr\ is marked by the blue circle, radiating from this point are 150 realisations of our proper motion constraints shown as coloured lines. The lines show the expected position of the SGR assuming linear motion and not accounting for the motion of the Sun. The lines are colour coded to indicate ages of 1--4, 4--7 and 7--10\,kyr of \sgr, as indicated by the colour bar. The low significance of the PM does allow for an origin of \sgr\ in most directions from its current position, albeit at a lower probability than a SSE origin. {\em Right:} A 1.28\,Ghz continuum image of \snr\ from the MeerKAT telescope. The current position of \sgr\ is again indicated by a blue circle, with the contours showing the \percentile{68} and \percentile{95} distribution of origins for the source based on our PM constraints, assuming a 16\,kyr age of the magnetar. A pink dashed circle guides the eye to the extent of \snr\ based on the morphology of the remnant to the North. The centre of this circle is indicated by the pink $\times$ symbol. The view of the left \hst\ panel is indicated by the red square.
}
\end{figure*}

The expected young age, and our relatively low PM value, can be used to determine that the present location of \sgr\ is close to its birth place. To visualise this, in Figure \ref{fig:sgr_origin_snr} (left) we plot random realisations of our PM, tracing back typical ages of magnetars. The origin is comfortably contained within our \hst\ imaging for typical magnetar ages.\footnote{Due to telescope orientation changes, the field of view for Epochs \epochone\ and \epochtwo\ are more significantly truncated along this region.} Searching for possible associations or clusters as the birthplace of the magnetar progenitor is, however, difficult given distance estimates to \sgr\ are significantly further than even state-of-the-art cluster searches allow \citep[e.g.][]{castroginard2020}. 

Instead, we turn our attention to the magnetar's spatial coincidence with \snr\ and reassess this in the light of our PM constraints. Similar analysis has been done for other SNR-related magnetars \citep[e.g.][]{tendulkar2013}. We show in Figure \ref{fig:sgr_origin_snr} (right) the position of the magnetar alongside contoured realisations of our PM traced back 16\,kyr. This nominal age is motivated by studies of the age of \snr\ \citep{ranasinghe2018, zhou2020}. The contours are overlaid on a radio continuum image at 1.28Ghz from the MeerKAT telescope \citep{meerkat2009}. A `by eye`-placed dashed circle indicates the extent of \snr\ based mainly on the bright northern lobe. When comparing the most likely origin of \sgr\ from the PM contours, we see that this matches up very well with the centre of the SNR (pink cross). Thus, although \sgr\ was already highly consistent with sharing a common origin with \snr, the PM constraints we derive here only serve to strengthen this association by suggesting the magnetars birth place (if it has an age equal to that of SNR) is almost exactly the geometric centre of \snr.

Searching for a potential surviving unbound binary companion to the magnetar is, in principle, possible.\footnote{Assuming that the NIR counterpart emission is not coming from a bound surviving companion, which is itself a distinct possibility (Section \ref{sec:niremission}).} However, the sheer density of sources in the field, the wide range of plausible PMs of the companion and the lack of colour information for comparably bright sources, makes drawing definitive conclusions tricky. Future characterisation of the field in order to build SEDs for the sources and compare with binary population synthesis models may allow for a more detailed study of any surviving companion.

\subsection{A comparison of magnetar and pulsar kick distributions}
\label{sec:compare_kick_distribution}

\begin{figure}
\includegraphics[width=\linewidth]{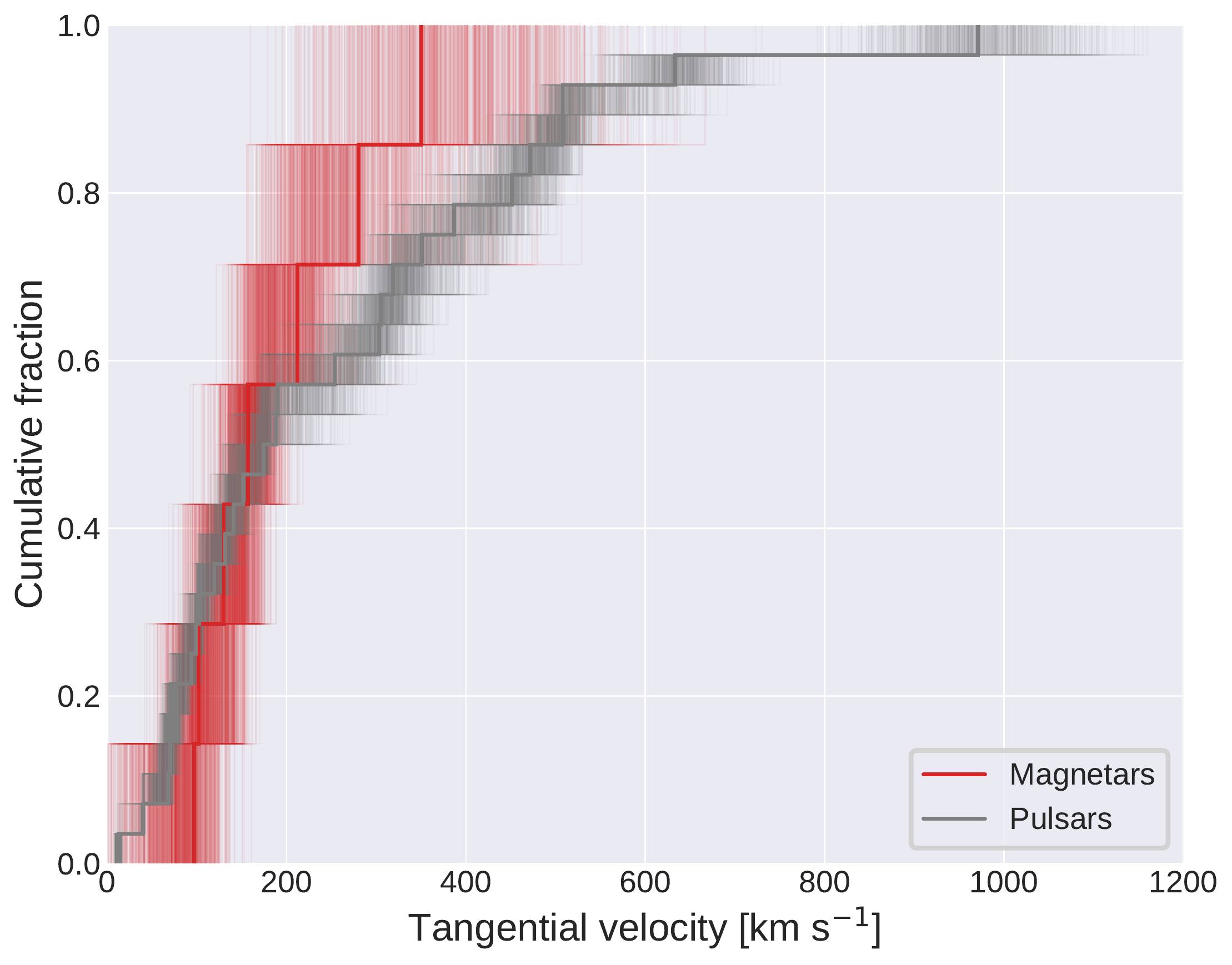}
\caption{
\label{fig:cumulative_vt}
The cumulative distribution of tangential velocities for magnetars and pulsars. Nominal values are presented as thicker lines, with 1000 realisations of each distribution, assuming Gaussian uncertainties, plotted as fainter lines. Performing Anderson-Darling tests between these 1000 realisations gave no results with a $>2\sigma$ significance, indicating there is currently no evidence to suggest magnetars velocities are distinct from the pulsar distribution. Data for the pulsars were taken from \citet{verbunt2017}, which have here been corrected for differential Galactic rotation and peculiar Solar motion using the prescription described in Section \ref{sec:absolute_astrometry_results}. Data for the magnetars were taken from this work and \citet{helfand2007, deller2012, tendulkar2012, tendulkar2013} -- these velocities were taken already corrected for Galactic rotation and Solar motion, using similar models but with different values. In practice, any difference in these corrections are dwarfed by the overall uncertainty on the values.
}
\end{figure}

There are currently few published PM constraints for magnetars. To our knowledge, the six presented in \citet{helfand2007, deller2012, tendulkar2012, tendulkar2013} are here added to by \sgr, giving a sample of seven. Adding our new value of $97\pm48$\kms\ to the values in table 7 of \citet{tendulkar2013}, we find the (mean, median and standard deviation) of magnetar tangential velocities are approximately (190, 160, 90)\kms. In Figure \ref{fig:cumulative_vt} we show a comparison of magnetar tangential velocities to the much more numerous ``young"" pulsars \citep{verbunt2017}. Given the close links between magnetars and pulsars, and even potential overlap in membership for some sources \citep[e.g.,][]{kaspi2005, rea2010}, we may expect their kick distributions to arise from a single distribution. Indeed, formally we found no evidence to reject this hypothesis, and conclude that they are indistinguishable, given current constraints. The lack of any high velocity magnetars (i.e. $\vtot \gtrsim 500$\kms), cf. the pulsar distribution, is not unusual given the comparative sample sizes.

\section{Summary}
\label{sec:summary}

We have presented new \hst\ NIR observations of \sgr\ that significantly extend the baseline of high-resolution observations for this magnetar to $\sim6$\,years. Using these we have constrained the tangential velocity of the source, finding a $\sim2.1\sigma$ detection of motion -- $\vtot=97\pm48$\kms\ -- that indicates it to be moving at a modest velocity compared to the distribution of pulsar kick velocities \citep{verbunt2017} and other magnetars. To properly assess the statistical significance of this velocity, additional constraints are required over a longer baseline (see discussion in Section \ref{sec:absolute_astrometry_results}). However, our conclusions remain largely-unchanged if we consider it as a null-detection, given our uncertainty on the measurement is low compared to typical magnetar velocities. The NIR brightness of the source has now been seen to vary by one magnitude, with it being significantly brighter in the most recent epoch, despite the NIR observation being taken during a period of X-ray quiescence. There is tentative evidence of shorter timescale variability in the NIR, although additional observations while the source is brighter are needed to confirm this. The origin of the NIR emission remains unclear, although indications of a lack of correlation with X-ray behaviour, if confirmed, would prove problematic for debris disk or magnetosphere models of the emission. Alternative origins, such as a binary companion, will require fuller characterisation of the NIR SED in order to be properly evaluated. Even with the increase in brightness, observations in the NIR of this still faint and crowded magnetar remain solely in the remit of space-based facilities such as \hst\ and James Webb Space Telescope.
Constraints on magnetar velocities remain sparse. With current statistics, they remain indistinguishable from the pulsar distribution.
\begin{acknowledgments}

JDL wishes to thank Rolly Bedin for fruitful discussions relating to \hst\ astrometry.
JDL acknowledges support from a UK Research and Innovation Future Leaders Fellowship (MR/T020784/1). AJL has received funding from the European Research Council (ERC) under the European Union’s Seventh Framework Programme (FP7-2007-2013) (Grant agreement No. 725246). 
KW acknowledges support through a UK Research and Innovation Future Leaders Fellowship (MR/T044136/1) awarded to dr.~B. Simmons.
The MeerKAT telescope is operated by the South African Radio Astronomy Observatory (SARAO), which is a facility of the National Research Foundation, an agency of the Department of Science and Innovation. We thank B. Stappers and I. Heywood for providing us the MeerKAT image of \snr. Based on observations made with the NASA/ESA Hubble Space Telescope, obtained at the Space Telescope Science Institute, which is operated by the Association of Universities for Research in Astronomy, Inc., under NASA contract NAS 5-26555. These observations are associated with programs 14055, 14502 and 16505.
\end{acknowledgments}

\vspace{5mm}
\facilities{HST (WFC3/IR), Swift, Fermi}

\software{Astropy \citep{astropy2013, astropy2018},
          DOLPHOT \citep{dolphot},
          lmfit \citep{lmfit},
          spalipy \citep{spalipy},
          stsynphot \citep{stsynphot},
          TOPCAT \citep{topcat},
          uncertainties\footnote{\url{http://pythonhosted.org/uncertainties/}}
          }




\bibliography{references}{}
\bibliographystyle{aasjournal}



\end{document}